\documentclass[12pt,preprint]{aastex}

\begin{document}


\title{Time delay of photons of different energies in multi-dimensional cosmological models}

\author{T. Harko$^1$ and K. S. Cheng$^2$}

\email{$^1$harko@hkucc.hku.hk, $^2$hrspksc@hkucc.hku.hk}

\affil{ Department of Physics, The University of Hong Kong,
Pokfulam Road, Hong Kong}


\date{May 4, 2004}

\begin{abstract}

We consider the general expressions for the time delay of photons
of different energies in the framework of multi-dimensional
cosmological models. In models with compactified extra-dimensions
(Kaluza-Klein type models), the main source of the photon time
delay is the time variation of the electromagnetic coupling, due
to dimensional reduction, which induces an energy-dependence of
the speed of light. A similar relation between the fine structure
constant and the multi-dimensional gauge couplings also appears in
models with large (non-compactified) extra-dimensions. For photons
of energies around 1 TeV, propagating on cosmological distances in
an expanding Universe, the time delay could range from a few
seconds in the case of Kaluza-Klein models to a few days for
models with large extra-dimensions. As a consequence of the
multi-dimensional effects, the intrinsic time profiles at the
emitter rest frame differ from the detected time profiles. The
formalism developed in the present paper allows the transformation
of the predicted light curves of various energy ranges of the
emitter into the frame of the observer, for comparison with
observations. Therefore the study of energy and redshift
dependence of the time delay of photons, emitted by astrophysical
sources at cosmological distances, could discriminate between the
different multi-dimensional models and/or quantum gravity effects.

\end{abstract}


\maketitle


\section{Introduction}

One of the most challenging issues of modern physics is the
existence of the extra-dimensions, idea proposed originally by
\citet{Kl19} and developed by \citet{Kl26}. Multi-dimensional
geometries are the natural framework for the modern string/M
theories \citep{Wi96} or brane models \citep{HoWi96}. String
models also provide a natural and self-consistent explanation for
the possible variation of the fundamental constants, as initially
suggested by \citet {Di38, Di39, Di79}. Hence the problem of the
extra-dimensions of the space-time continuum is closely related to
the problem of the variations of the fundamental constants, like,
for example the fine structure constant or the speed of light (for
a recent review of the experimental and theoretical studies and
the present status of these fields see \citet{Uz03} and \citet
{Ma03}. Most of the theories with extra-dimensions contain a
built-in mechanism, which allows the variation of the fundamental
constants. Within the multi-dimensional approach the physical
interactions are described by a theory formulated in $4+D$
dimensions, and the conventional four-dimensional theory appears
as a result of a process of dimensional reduction. Couplings in
four dimensions are determined by a set of constants of the
multidimensional theory and the size $A$ of the space of
extra-dimensions. The multi-dimensional constants are assumed to
be genuinely fundamental and, consequently, they do not vary with
time. On the other hand it is natural to assume that in an
astrophysical or cosmological context $A$ varies with time,
similarly to the scale factor $a$ of our four-dimensional
Universe. But this leads to the time variation in four dimensions
of the fundamental constants, like fine structure constant $\alpha
$ or the gravitational coupling $G$. Moreover, since their time
dependence is given by the same factor $A$, the time variations of
$\alpha $ and $G$ could be correlated \citep{La02}.

The search for a unification of quantum mechanics and gravity is
likely to require a drastic modification of the present day
deterministic representation of the space-time properties. There
is at present no complete mathematical model for quantum gravity,
and no one of the many different models proposed so far can give a
satisfactory description of the physics on characteristic scales
near the Planck length $l_{P}$. However, in several of the
approaches trying to find a theory of quantum gravity the vacuum
can acquire non-trivial optical properties, because of the
gravitational recoil effects induced by the motion of the
energetic particles. The recoil effects may induce a non-trivial
refractive index, with photons at different energies travelling at
different velocities \citep{El00c}. Photon
polarization in a quantum space-time may also induce birefringence \citep{Ga99}%
, while stochastic effects in the vacuum could give rise to an
energy dependent diffusive spread in the velocities of different
photons \citep{El00, E101}.

Therefore a large class of physical models, incorporating quantum
gravitation and/or multi-dimensional field theories predict that
the propagation of particle in vacuum is modified due to the
supplementary effects induced by the modification of the standard
general relativity. In particular, the possible violation of the
Lorentz invariance or the existence of extra-dimensions can be
investigated by studying the propagation of high energy photons
emitted by distant astrophysical sources \citep{Am98}.

The highest energy extra-galactic $\gamma $-ray sources in the
known universe are the active galaxies called blazars: objects
that emit jets of relativistic plasma aimed directly at us.
Objects known as high frequency BL Lac objects (HBLs) are expected
to emit photons in the multi-TeV energy range. Only the nearest
ones are expected to be observable in TeV energies, the others
being hidden by intergalactic absorption \citep{St96}.

Extragalactic photons with the highest energy yet observed
originated in a powerful flare coming from the giant elliptical
active galaxy known as
Markarian 501 in 1997 \citep{Ah97} and from Markarian 421 in 2001 \citep{Kr01}%
. The high energy flux of these emissions permitted detailed
spectra to be extracted. Since as many as $25,000$ photons were
detected, the spectra were derived with a high statistical
accuracy. The observation of the TeV photons from the Markarian
501 and 421 allows to impose some constraints on the quantum
gravity scale and on the breaking of the Lorentz invariance
\citep{Bi, St03}. Quantum gravity phenomena are a result of the
quantum fluctuations on the Planck scale $M_{P}=\sqrt{\hbar c/G}$.
In models involving large extra-dimensions, like, for example in
the brane-world models \citep{Ra98,Ra98a}
the energy scale at which gravity becomes strong can be much smaller than $%
M_{P}$, with the quantum gravity scale $M_{QG}$ approaching the TeV scale.

The data from the TeV gamma -ray flare associated with the active
galaxy Markarian 421 have been used to place bounds on the
possible energy dependence of the speed of light in the context of
an effective quantum gravitational energy scale in \citep{Bi}. The
limits derived indicate this energy scale to be higher than $6
\times 10^{16}$ GeV for the approach to quantum gravity in the
context of D-brane string theory. By assuming a modified
dispersion relation between the energy $E_{\gamma }$ and the
momentum $p_{\gamma }$ of the photon of the form $E_{\gamma
}^{2}=p_{\gamma }^{2}-p_{\gamma }^{3}/M_{QG}$, and a similar
relation for the electron, one can obtain the constraint
$M_{QG}\geq E_{\gamma }^{3}/8m_{e}^{2}$ for the quantum gravity
scale \citep{St03}. Since pair
production occurs for energies of at least $20$ TeV, it follows that $%
M_{QG}\geq 0.3M_{P}$. The results also indicate an absence of evidence for
the violation of the Lorentz invariance, as proposed by some quantum gravity
and multi-dimensional models.

Strong constraints on Lorentz violating microscopic structures of
space-time, like, for example, discreteness, non-commutativity, or
extra dimensions can be obtained from the observation of $100$ MeV
synchrotron radiation from the Crab nebula \citep{Ja03}. The Crab
synchrotron emission has been observed to extend at least up to
energies of $100$ MeV, just before the Compton hump begins to
contribute to the spectrum \citep{At96, Hi98}. The magnetic field
in the emission region has been estimated to a value between
$0.15-0.6$ mG \citep{Hi98}. To produce the observed radiation of
$100$ MeV in this field requires a relativistic $\gamma $-factor
of the order of $\gamma =3\times 10^{9}$, corresponding to an
electron energy of the order of $1500$ TeV, with an electron
velocity differing from $c$ by less then $10^{-19}c$. Then the
observation of 100 MeV
synchrotron radiation from the Crab nebula gives the constraint $%
E_{QG}>10^{26}$ GeV \citep{Ja03}. Hence this observation rules out
this type of Lorentz violation, providing an important constraint
on theories of quantum gravity and imposes a stringent constraint
on any modification of the dispersion relations of the electron
that might be induced by quantum gravity.

The spectrum of the synchrotron radiation from the coupling of an
electrically-charged particle to an external magnetic field, in
the presence of quantum-gravity effects of the general form
$\left( E/M_{QG}\right) ^{\alpha }$ has been derived in
\citet{ElM03}. The synchrotron constraint from the Crab Nebula
practically excludes $\alpha \sim 1.74$ for $M_{QG}\sim M_{P}\sim
1.2\times 10^{19}$ GeV. The model suggests a linear modification
of the dispersion relation for the photon, but not for the
electron, and hence is compatible with known constraints from the
Crab Nebula. New constraints on possible Lorentz symmetry
violation of the order of $E/M_{P}$ for electrons and photons in
the framework of effective field theory, by using the absence of
vacuum birefringence in the recently observed polarization of MeV
emission from a gamma ray burst, and the absence of vacuum
Cerenkov radiation from the synchrotron electrons in the Crab
nebula, have been derived in \citet{Jal03}. These constraints
allow to improve the previous bounds by eleven and four orders of
magnitude, respectively.

The possibility of the use of the high energy radiation from
gamma-ray pulsars to place limits on quantum gravity effects has
been suggested by \citet{Ka99}. The emission from the Crab pulsar
at energies above $2$ GeV trails that at $70-100$ MeV by no more
than $0.35$ ms ($95\%$ confidence). This effect places a lower
bound on the energy scale of quantum gravitational effects on the
speed of light of $1.8\times 10^{15}$ GeV. In the near future this
bound might be improved by two orders of magnitude by
observation of pulsations from the Crab at higher energies, of the order of $%
50-100$ GeV.

The confirmation that at least some gamma-ray bursts (GRBs) are
indeed at cosmological distances raises the possibility that
observations of these could provide interesting constraints on the
fundamental laws of physics (for recent reviews on GRBs see
\citep{ZhMe03, ChLu01}). The fine-scale time structure and hard
spectra of GRB emissions are very sensitive to the possible
dispersion of electromagnetic waves in vacuo, with velocity
differences $\Delta u\sim E/E_{QG}$, as suggested in some
approaches to quantum gravity. GRB measurements might be sensitive
to a dispersion scale $E_{QG}$ comparable to the Planck energy
scale $E_{P}\sim 10^{19}$ GeV, sufficient to test some of these
theories \citep{Am98}. Hence the study of short-duration photon
bursts propagating over cosmological distances is the most
promising way to probe the quantum gravitational and/or the
effects related to the existence of extra-dimensions. The
modification of the group velocity of the photons by the quantum
effects would affect the simultaneity of the arrival times of
photons with different energies. Thus, given a distant, transient
source of photons, one could measure the differences in the
arrival times of sharp transitions in the signals in different
energy bands. A key issue in such a probe is to distinguish the
effects of the quantum-gravity/multi-dimensional medium from any
intrinsic delay in the emission of particles of different energies
by the source. The quantum-gravity effects should increase with
the redshift of the source, whereas source effects would be
independent of the redshift in the absence of any cosmological
evolution effects. Therefore it is preferable to use transient
sources with a known spread in the redshift $z$. The best way to
probe the time lags that might arise from quantum gravity effects
is to use GRBs with known redshifts, which range up to $z\sim 5$.

Data on GRBs may be used to set limits on variations in the
velocity of light. This has been illustrated, by using BATSE and
OSSE observations of the GRBs that have recently been identified
optically, and for which precise redshifts are available, in
\citet{Ma00}.  A regression analysis can be performed to look for
an energy-dependent effect that should correlate with redshift.
The analysis of GRBs data yield a limit $M_{QG}\sim 10^{15}$ GeV
for the quantum gravity scale. The study of the the times of
flight of radiation from gamma-ray bursts with known redshifts has
been considerably improved by using a wavelet shrinkage procedure
for noise removal and a wavelet `zoom' technique to define with
high accuracy the timings of sharp transitions in GRB light curves
\citep{El03}. This procedure optimizes the sensitivity of
experimental probes of any energy dependence of the velocity of
light. These wavelet techniques have been applied to $64$ ms and
TTE data from BATSE, and also to OSSE data. A search for time lags
between sharp transients in GRBs light curves in different energy
bands yields the lower limit $M_{QG}\geq 6.9\times 10^{15}$GeV on
the quantum-gravity scale in any model with a linear dependence of
the velocity of light, $c\sim E/M_{QG}$.

High energy GeV emissions from GRBs have already been detected
\citep{So94, Hu94}. There is also tentative evidence for TeV
emissions \citep{At00, Po03}. The production of TeV photons is
also predicted by most of the GRB theories. The emission
mechanisms for TeV photons include electron inverse Compton
emission and synchrotron emission from the protons accelerated by
GRB shocks. The shocks could be internal shocks, external forward
shocks or external reverse shocks of GRBs. Such very high energy
photons at cosmological distances may largely be absorbed by
interacting with the cosmic infrared background radiation
\citep{Ma96}.

There are many astrophysical mechanisms that could produce a delay
in the arrival time of high energy photons or cosmic rays. The
electron inverse scattering of the created electron-positron pairs
off the cosmic microwave background photons will produce delayed
MeV-GeV emission. There are two likely mechanisms causing the time
delay. One is the angular spreading effect of the secondary pairs,
that is, the scattered microwave photons deviate from the
direction of the original TeV photons by an angle $\sim 1/\gamma
$, where $\gamma $ is the Lorentz factor of the electron-positron
pairs \citep{Ch96, Da02, WaCh03}. Another mechanism is related to
the deflection of the direction of propagation of the pairs in the
intergalactic magnetic field, if this field is sufficiently strong
\citep{Pl95}. It is important to note that all these delay
mechanisms predict that the low energy photon come late. But in
the models with extra-dimensions the high energy photon comes
later.

It is the purpose of the present paper to point out some other
possible sources of time delay for high energy photons from GRB
emissions. Namely, we shall consider the effects of the existence
of the extra-dimensions on the propagation of high energy photons
in two distinct physical scenarios: Kaluza-Klein theories with
compactified extra-dimensions and models with large
extra-dimensions. In Kaluza-Klein theories the time variation of
the scale factor of the space of extra dimensions leads to a time
variation of both the gravitational constant $G$ and of the fine
structure constant $\alpha $. The variation of $\alpha $ is
photon-energy dependent, and via the time delay allows the
estimation of the size of the (compactified) extra-dimension. In
the case of the large (non-compactified) extra-dimensions, there
is no multi-dimensional physical mechanism to induce a variation
of $\alpha $. Recently, using high resolution spectroscopy of QSO
absorption spectra, a time variation of $\alpha $ has been
reported \citep{We1, We2, We3}. The detected rate of change of
$\alpha $ is of the order $\Delta \alpha /\alpha \sim -10^{-5}$ at
a redshift $z\sim 1.5$.

The present paper is organized as follows. In Section II we
discuss the most important physical processes that could lead to
the variation of the electromagnetic coupling, in the different
versions of the multi-dimensional models. The basic equations for
the time delay of photons in Kaluza-Klein and Randall-Sundrum type
models are obtained in Section III. In Section IV we discuss and
conclude our results.

\section{Variation of the electromagnetic coupling in models with extra
dimensions }

The starting point in the multi-dimensional approach of describing the
fundamental interactions by Kaluza and Klein  is the consideration of the
pure Einstein gravity in a multi-dimensional space-time $M^{4}\times K^{D}$,
described by the multi-dimensional metric tensor $\hat{g}_{MN}$. Here $M^{4}$
is the four-dimensional space-time and $K^{D}$ is the compact manifold of
extra-dimensions. Generally the reduced theory contains the Einstein gravity
and the Yang-Mills fields with the gauge group determined by the isometry
group of the space of extra-dimensions. The action of the multi-dimensional
Kaluza-Klein theory is the action for the pure Einstein theory on $%
M^{4}\times K^{D}$, with the action given by \citep{Ov97}
\begin{equation}
S=\frac{1}{16\pi G_{\left( 4+D\right) }}\int d^{4+D}\hat{x}\sqrt{-\hat{g}}%
R^{\left( 4+D\right) },
\end{equation}
where $\hat{g}=\det \left( \hat{g}_{MN}\right) $, $R^{\left( 4+D\right) }$
is the scalar curvature in $M^{4}\times K^{D}$ and $G_{\left( 4+D\right) }$
is the multi-dimensional gravitational constant, which is assumed to be a
true constant and does not depend on time. According to the procedure of
dimensional reduction, the $\left( \mu ,\nu =0,1,2,3\right) $ components of
the metric tensor are identified as the four-dimensional metric tensor,
while certain combinations of the rest of the components are identified as
gauge field multiplets $A_{\mu }$ and scalar fields $\phi _{m},m=1,2...$.
After the mode expansion of these fields, with the coefficients of the
expansion (interpreted as four-dimensional fields) depending only on $x^{\mu
},\mu =0,1,2,3$ we obtain the four-dimensional action
\begin{equation}
S_{0}=\int d^{4}x\left[ \frac{1}{16\pi G(t)}R^{(4)}+\sum_{i}\frac{1}{%
4g_{i}^{2}(t)}Tr\left( F_{\mu \nu }^{(i)}F^{i(\mu \nu )}\right) \right] ,
\label{eq2}
\end{equation}
where $G(t)\equiv G_{4}(t)$ is the four-dimensional gravitational
constant. In obtaining Eq. (\ref{eq2}) we have considered only
zero modes of the mode expansion. The parameters $g_{i}\left(
t\right) $ are the gauge couplings, and the index $i$ labels the
simple subgroups of the gauge group. The general reduction of the
initial Kaluza-Klein action $S$ also gives terms including
non-zero modes of the gravitational, gauge and scalar fields
\citep{Ov97}. The scalar fields give highly non-linear interaction
terms and are coupled non-minimally to the gravitational and gauge
fields. In the following we neglect, for simplicity, the
contribution of these scalar
fields.

Identifying the gravitational and gauge couplings from the action $%
S_{0}$ for the zero modes one obtains the following expressions
for $G(t)$ and $g_{i}^{2}(t)$ in terms of $G_{\left( 4+D\right) }$
and the radius $\Phi \left( t\right) $ of the space of the
extra-dimensions \citep{Lo03}:
\begin{equation}
G\left( t\right) =\frac{G_{\left( 4+D\right) }}{V_{D}\left( t\right) }%
,g_{i}^{2}(t)=\tilde{k}_{i}\frac{G_{\left( 4+D\right) }}{\Phi ^{2}\left(
t\right) V_{D}\left( t\right) },
\end{equation}
where $V_{D}\left( t\right) \sim \Phi ^{D}\left( t\right) $ is the
volume of the space of the extra-dimensions and $\tilde{k}_{i}$
are coefficients which depend on the isometry group of $K^{D}$.
The fine structure constant $\alpha \left( t\right) $ is given by
a linear combinations of $g_{i}^{2}(t)$, with the specific
relation depending on the gauge group and the scheme of
spontaneous symmetry breaking. Generally,
\begin{equation}
\alpha \left( t\right) =k_{1}\frac{G_{\left( 4+D\right) }}{\Phi ^{2}\left(
t\right) V_{D}\left( t\right) },
\end{equation}
where $k_{1}$ is a constant. Since $\dot{V}_{D}\left( t\right)
/V_{D}\left( t\right) =d\left( \dot{\Phi}/\Phi \right) $, for the
time variation of the fine structure constant we obtain
\begin{equation}
\frac{\dot{\alpha}}{\alpha }=-\left( D+2\right) \frac{\dot{\Phi}}{\Phi }.
\end{equation}

The variation of the fine structure constant in a more general model, in
which a Yang-Mills type field is also included in the $\left( 4+D\right) $%
-dimensional space-time was also considered in \citet{Lo03}. The
action is
\begin{equation}
S=\int d^{4+D}\hat{x}\sqrt{-\hat{g}}\left[ \frac{1}{16\pi G_{\left(
4+D\right) }}R^{\left( 4+D\right) }+\frac{1}{4g_{\left( 4+D\right) }^{2}}%
Tr\left( \hat{F}^{MN}\hat{F}_{MN}\right) \right] ,  \label{ym}
\end{equation}
with $g_{\left( 4+D\right) }^{2}$ the multi-dimensional coupling,
supposed to be constant in time. In this case the dimensionally
reduced theory includes the Einstein gravity and the
four-dimensional gauge fields plus scalar fields with a quartic
potential. The four dimensional gravitational and fine structure
constants are given by
\begin{equation}
G(t)=\frac{G_{\left( 4+D\right) }}{V_{D}\left( t\right) },\alpha \left(
t\right) =k_{2}\frac{G_{\left( 4+D\right) }}{V_{D}\left( t\right) },
\end{equation}
where $k_{2}$ is a constant. The time variation of the fine-structure
constant is
\begin{equation}
\frac{\dot{\alpha}}{\alpha }=-D\frac{\dot{\Phi}}{\Phi }.
\end{equation}

We consider now models with large extra-dimensions (brane world
models), as initially considered in \citet{Ra98}. In the $5D$
space-time the brane-world is located at $Y(X^{I})=0$, where
$X^{I},\,I=0,1,2,3,4$, are 5-dimensional coordinates. The
effective action in five dimensions is \citep{MW00, Chnu}
\begin{equation}
S=\int d^{5}X\sqrt{-g_{5}}\left( \frac{1}{2k_{5}^{2}}R_{5}-\Lambda
_{5}\right) +\int_{Y=0}d^{4}x\sqrt{-g}\left(
\frac{1}{k_{5}^{2}}K^{\pm }-\lambda +L^{\textrm{matter}}\right),
\end{equation}
where $k_{5}^{2}=8\pi G_{5}$ the 5-dimensional gravitational
coupling constant, $\Lambda _{5}$ is the cosmological constant in
the bulk. $x^{\mu },\,\mu =0,1,2,3$, are the induced 4-dimensional
brane world coordinates. $R_{5}$ is the 5D intrinsic curvature in
the bulk and $K^{\pm }$ is the extrinsic curvature on either side
of the brane.

Assuming a metric of the form
$ds^{2}=(n_{I}n_{J}+g_{IJ})dx^{I}dx^{J}$, with $n_{I}dx^{I}=d\chi
$ the unit normal to the $\chi =\textrm{constant}$ hypersurfaces
and $g_{IJ}$ the induced metric on $\chi =\textrm{constant}$
hypersurfaces, the effective four-dimensional gravitational
equations on the brane take the form \citep{SMS00, SaShMa00, Chph,
Ch2}:
\begin{equation}
G_{\mu \nu }=-\Lambda g_{\mu \nu }+k_{4}^{2}T_{\mu \nu }+k_{5}^{4}S_{\mu \nu
}-E_{\mu \nu },  \label{EqEinstein}
\end{equation}
where
\begin{equation}
S_{\mu \nu }=\frac{1}{12}TT_{\mu \nu }-\frac{1}{4}T_{\mu }{}^{\alpha }T_{\nu
\alpha }+\frac{1}{24}g_{\mu \nu }\left( 3T^{\alpha \beta }T_{\alpha \beta
}-T^{2}\right) ,
\end{equation}
and $\Lambda =k_{5}^{2}(\Lambda _{5}+k_{5}^{2}\lambda
^{2}/6)/2,\,k_{4}^{2}=k_{5}^{4}\lambda /6$.  $E_{IJ}=C_{IAJB}n^{A}n^{B}$. $%
C_{IAJB}$ is the 5-dimensional Weyl tensor in the bulk and $\lambda $ is the
vacuum energy on the brane. $T_{\mu \nu }$ is the matter energy-momentum
tensor on the brane and $T=T^{\mu }{}_{\mu }$ is the trace of the
energy-momentum tensor.

The reduction formula expressing the four-dimensional Planck mass
$M_{Pl}$ in terms of the fundamental (five-dimensional) mass scale
$M=\left( 16\pi \hat{G}_{(5)}\right) ^{-1/3}$ $\sim k$ has been
derived in \citet{Bo02}.  The result is
\begin{equation}\label{Pl}
M_{Pl}^{2}=\frac{M^{3}}{k}\left( e^{2\pi k\Phi }-1\right) .
\end{equation}

Since in the models with large extra-dimensions all matter fields
are localized on the brane and do not depend on the radius of the
extra-dimension, there is no simple mechanism for the variation of
the fine structure constant. A possible form of the variation of
$\alpha $ has been obtained in \citet{Lo03}, giving
\begin{equation}\label{bulk}
\alpha \left( t\right) =k_{3}\frac{g_{(5)}^{2}}{\Phi \left( t\right) },\frac{%
\dot{\alpha}}{\alpha }=-\frac{\dot{\Phi}}{\Phi },
\end{equation}
where $k_{3}$ is a constant.

Finally, we shall briefly consider the problem of the quantum
corrections to the expressions for the fine structure constant
presented above. This question has been discussed in \citet{La02}
and \citet{Lo03}. To relate the value of $\alpha $ obtained at the
scale $M_{\Phi }=\Phi ^{-1}\left( t\right) $ to its value at some
low energy scale $\mu $ by taking into account quantum corrections
one must use the renormalization group formulas for running
couplings, which gives for the time variation of $\alpha $
\begin{equation}
\frac{\dot{\alpha}\left( \mu ,t\right) }{\alpha \left( \mu ,t\right) }=\frac{%
\dot{\alpha}\left( M_{\Phi },t\right) }{\alpha \left( M_{\Phi },t\right) }%
-\alpha \left( \mu ,t\right) A\frac{\dot{\Phi}}{\Phi }\left( 1+\ln
\mu \Phi \right) ,  \label{eqq}
\end{equation}
where $A$ is a constant of order one.

The second term in Eq. (\ref{eqq}) is of order $O\left( \alpha \right) $.
Hence it is a subdominant term, thus showing that quantum effects does not
significantly affect the time variation of the fine structure constant.

\section{Time delay of photons in multi-dimensional expanding Universes}

We consider first the propagation of gamma-rays from GRBs in the
Kaluza-Klein type models. For simplicity we restrict our
discussion to the five-dimensional case. Hence we assume a flat
Friedmann-Robertson-Walker type background metric of the form
\begin{equation}
ds^{2}=c^{2}dt^{2}-a^{2}\left( t\right) \left[ dr^{2}+r^{2}\left( d\theta
^{2}+\sin ^{2}\theta d\varphi ^{2}\right) \right] +\varepsilon \Phi
^{2}\left( t\right) dv^{2},
\end{equation}
where $a$ is the scale factor of the Universe, $\varepsilon =\pm 1$ and $%
\Phi $ is the scale factor of the fifth dimension, denoted by $v$. We also
assume that the time variation of the fine-structure constant is entirely
due to the change in the speed of light $c$. Therefore we neglect any
possible time variation of the electric charge or Planck's constant. Then
the time variation of the speed of light can be related to the size of the
fifth dimension by means of the general equation
\begin{equation}\label{c}
\frac{\Delta \dot{c}}{\Delta c}=\beta \varepsilon
\frac{\dot{\Phi}}{\Phi },
\end{equation}
where $\beta =1$ in the case of the Einstein-Yang-Mills model and $\beta =3$
for the case of the pure Einstein gravity in five dimensions. Eq. (\ref{c})
can be integrated to give
\begin{equation}
c=c_{0}\left(1+\varepsilon\Phi ^{\beta }\right),
\end{equation}
where $c_{0}$ is an arbitrary integration constant. In order to
find a simple and directly testable relation between the radius of
the extra-dimension and the energy of the photon, we shall assume,
following the initial proposal in \citet{Ma90, Ma90a} that the
mass of a body (and the associated energy) corresponds to the
length of a ''line segment'' of the fifth subspace. Such a
relation embodies the spirit of the Mach's principle in the sense
that the inertial mass depends on the distribution of the matter
in the Universe. In a more general formulation, we shall assume
that the variables parameters $c$, $G$ and the photon energy
$E=h\nu $ are related to the metric tensor component of the fifth
dimension by means of the equation \citep{MaHa99}
\begin{equation}
\frac{G\left( t\right) E}{c^{4}}=\frac{1}{\gamma }\int_{v^{0}}^{v}\sqrt{\left| g_{44}\right| }%
dv=\frac{1}{\gamma }\int_{v^{0}}^{v}\Phi dv,  \label{mak}
\end{equation}
where $\gamma $ is an arbitrary constant.

If $\Phi $ is independent of $v$, as is the case in models with
compactified extra-dimensions, Eq. (\ref{mak}) gives
\begin{equation}
\Phi =\frac{\gamma GE}{c^{4}\left( v-v^{0}\right)
}=\frac{E}{E_{K}},
\end{equation}
where we denoted $E_{K}=c^{4}\Delta v/\gamma G$, with $\Delta
v=v-v^{0}$ describing the variation of the size of the fifth
dimension between the moments of the emission and detection of a
photon. Therefore the energy-dependence of the speed of light of
the photon is given by
\begin{equation}
c=c_{0}\left[1+\varepsilon \left( \frac{E}{E_{K}}\right) ^{\beta
}\right]. \label{eq7}
\end{equation}

We consider two photons emitted during a gamma-ray burst with
present day energies $E_{1}$ and $E_{2}$. At earlier epochs, their
energies would have been blue-shifted by a factor $1+z$. Then it
follows that the difference in the velocities of the two photons
is given by
\begin{equation}
\Delta c=c_{0}\frac{\Delta E\left( 1+z\right) }{E_{K}},\beta =1,
\end{equation}
and
\begin{equation}
\Delta c=c_{0}\frac{f\left( E_{1},E_{2}\right) \left( 1+z\right) ^{3}}{%
E_{K}^{3}},\beta =3,
\end{equation}
respectively, where we denoted $\Delta E=E_{1}-E_{2}$ and $f\left(
E_{1},E_{2}\right) =E_{1}^{3}-E_{2}^{3}$. A linear energy
dependence of the difference of the photon velocities has also
been considered in \citep{El03} as a result of the
dispersion-relation analysis of the Maxwell equations in the
non-trivial background metric perturbed by the recoil of a massive
space-time defect during the scattering of a low energy photon or
neutrino.

For light propagating from cosmological distances the differential
relation between time and redshift is \citep{El03}
\begin{equation}
dt=-H_{0}^{-1}\frac{dz}{\left( 1+z\right) h(z)},
\end{equation}
where
\begin{equation}
g\left( z\right) =\sqrt{\Omega _{\Lambda }+\Omega _{M}\left( 1+z\right) ^{3}}%
,  \label{h}
\end{equation}
$H_{0}=72$ km s$^{-1}$ Mpc$^{-1}$ \citep{Fr01}, $\Omega _{M}\approx 0.3$ and $%
\Omega _{\Lambda }\approx 0.7$ are the mass density parameter and
the dark energy parameter, respectively \citep{Pe03}. A particle
with a velocity $c$ travels an elementary distance
$cdt=-H_{0}^{-1}cdz/\left( 1+z\right) g(z)$, with the difference
$\Delta L$ in the distances covered by the two particles given by
$\Delta L=H_{0}^{-1}\int_{0}^{z}\Delta cdz/\left( 1+z\right) g(z)$
\citep{El03}. By taking into account the expression for $\Delta c$
we obtain the following equations describing the time delay of two
photons:
\begin{equation}
\Delta t=H_{0}^{-1}\frac{\Delta E}{E_{K}}\int_{0}^{z}\frac{dz}{g(z)}%
,\beta =1,  \label{b1}
\end{equation}
\begin{equation}
\Delta t=H_{0}^{-1}\left[ \frac{f\left( E_{1},E_{2}\right) }{E_{K}}%
\right] ^{3}\int_{0}^{z}\frac{\left( 1+z\right) ^{2}dz}{g(z)},\beta =3.
\label{b3}
\end{equation}

In the case of isotropic homogeneous cosmological models with
large extra-dimensions there is a non-zero contribution from the
$5$-dimensional Weyl tensor from the bulk, expressed by a scalar
term $U$, called dark radiation \citep{Chnu, Ha03, Ch03}. The
``dark radiation'' term is a pure bulk (five dimensional) effect,
therefore we cannot determine its expression without solving the
complete system of field equations in $5$ dimensions. In the case
of a Friedmann-Roberston-Walker type cosmological model the
expression of the dark radiation is \citep{Chph, Da02a}
\begin{equation}
U=\frac{U_{0}}{a^{4}},
\end{equation}
with $U_{0}$ an arbitrary constant of integration. Since the fifth
dimension is large, the scale factor $\Phi $ can also be a
function of $v$. Hence an explicit knowledge of the $v$-dependence
of $\Phi $ is needed in order to derive the speed of light- photon
energy dependence. However, taking into account Eq. (\ref{bulk}),
which shows a linear dependence of $\alpha $ on the scale of the
fifth dimension, we can assume that the time delay between two
different energy photons emitted during a gamma ray burst is given
by
\begin{equation}
\Delta t=H_{0}^{-1}\frac{\Delta E}{E_{F}}\int_{0}^{z}\frac{dz}{h(z)},
\end{equation}
where $E_{F}$ is the energy scale associated with the large
extra-dimensions and
\begin{equation}
h(z)=\sqrt{\Omega _{\Lambda }+\Omega _{M}\left( 1+z\right) ^{3}+\Omega
_{U}\left( 1+z\right) ^{4}},
\end{equation}
with $\Omega _{U}$ the dark radiation parameter.

\section{Discussions and final remarks }

In order to calculate the delay in the gamma ray photons arrival
time in the Kaluza-Klein type models we need to estimate first the
Kaluza-Klein energy scale $E_{K}=c^{4}\Delta v/\gamma G$ for which
the effects from the extra-dimensions become important. Assuming
that the size and the variation of the extra-dimension $\Delta v$
is of the same order as the Planck length, $\Delta
v=l_{P}=1.6\times 10^{-33}$ cm, the Kaluza-Klein energy scale is
equal, for $\gamma =1$, to $E_{K}=1.2\times 10^{19}$ GeV. Of
course, a large value of $\gamma $ can decrease the Kaluza-Klein
energy scale. The variation of the difference in the photon
arrival time, for different values of the photon energy is
presented, for the Einstein-Yang-Mills model corresponding to
$\beta =1$ and with $\gamma =1$, in Fig. 1.

Due to the power three energy dependence of the time delay in the
pure Einstein gravity model ($\beta =3$) the value of $\Delta t$
is extremely small for $\gamma =1$, corresponding to the
$E_{K}=1.2\times 10^{19}$ GeV energy scale. In this case $\Delta
t\approx 10^{-30}$ s even for photon energies of the order of $1$
TeV. In order to obtain some observable effects a very large value
of $\gamma $, of the order of $\gamma =10^{10}$ is required. The
variation of the photon time delay in the case of the Kaluza-Klein
compactified model with pure Einstein gravity is represented in
Fig. 2.

The analysis of the BATSE and OSSE data has imposed a lower limit
$E_{QG}\geq 6.9\times 10^{15}$ GeV on the quantum gravity scale in
the linear model \citep{El03}, which is much smaller than the
Kaluza-Klein energy scale we have considered. If the fundamental
energy scale is of the order of $E_{QG}$, then the time delay
between TeV and KeV/MeV photons could have larger values than
those considered in the present approach.

To generate the correct hierarchy between the Planck scale and the
TeV scale in models with large extra-dimensions, the product
$k\Phi $ must be of the order of $k\Phi \approx 11-12$
\citep{Bo02}. There are no simple mechanisms to describe the
temporal evolution of the fine structure constant in this type of
models, a variation of $\alpha $ requiring the consideration of
bulk gauge and, perhaps, fermionic fields. However, based on the
analogy with the Kaluza-Klein case one can assume a linear
dependence of the time delay on $\Delta E$, with the
characteristic energy scale $E_F$ a parameter to be determined
from observations. The variation of the time delay for two
different energy photons is represented, for $E_F=7\times 10^{15}$
GeV, in Fig. 3.

In deriving the equations for the time delay due to the presence
of the effects of the extra-dimensions we have used a crucial
assumption, namely, we have considered that there is an intrinsic
synchronization of pulsing emission of photons at different energy
ranges. This assumption allows the consistent determination, by
comparison with observations, of the parameters determining the
effects of the extra-dimensions on the photon propagation.
However, if the emission of the photons of different energies is
not synchronized, the simple processing of data for the delay of
the photons is not enough to constrain the multi-dimensional
effects. There are many proposed explanations, which could produce
a delayed time scale at the emitter between, for example, GeV and
keV/MeV photons, as observed in the case of GRB 940217
\citep{Hu94}. These explanations include interaction of TeV
photons with cosmic infrared background photons \citep{Pl95},
interactions of ultrarelativistic protons with a dense cloud
\citep{Ka94} or inverse Compton scattering in early forward and
reverse shocks \citep{MeRe94}. In these cases, in order to
consistently determine the time delay between photons at different
energies, due to multi-dimensional propagation effects, the
knowledge of the initial time profiles of MeV, GeV and TeV photons
is also required. The description of the non-synchronized emission
time profiles is also model-dependent, and thus in this case it is
difficult to clearly distinguish between emission and propagation
effects. However, despite all these possibilities of producing a
time delay of different energy photons at the emitter, in our
model we make two very clear predictions, namely, (1) that the
high energy (TeV) photons have the higher delay and (2) that the
delay time scales are correlated with the redshift $z$ of the
emitting source, because the delay results from propagation
effects. In the future, if enough observational data will be
available, the study of the energy and redshift dependence of the
delay, $\Delta t=\Delta t\left(E_{\gamma } ,z\right)$ could lead
to the possibility of discriminating between emission and
propagation effects.

We also suggest that if there is an intrinsic synchronization of
the photon emission at different energies, then due to the effects
of the extra-dimensions the detected time profiles between the
KeV/MeV/GeV and the TeV bursts should be very different. We want
to emphasize that it is better to measure the time profile
difference between KeV/MeV and TeV photons, instead of the
difference between KeV/MeV and GeV photons, in order to avoid the
contamination or other effects. For an emitter with an initial
Gaussian time profile, $e^{-\left(t/\tau \right)^2}$, where $\tau$
is the duration of the burst, the pulse shape of the $1-10$ TeV
photons is shown in Fig. 4. We have considered that the GRB
occurred at a redshift of $z=3$, and considered the initial
duration of the burst $\tau =1$ s. For the time delay of the
photons we have adopted the linear model, with the fundamental
energy scale of the same order of magnitude as the Planck scale,
$E_K=1.2\times 10^{19}$ GeV. In this case the time delays of the
TeV photons with respect to the KeV/MeV/GeV photons are $\Delta
t_1=52.9$ s for the $1$ TeV photon and $\Delta t_2=529.3$ s for
the $10$ TeV photon, respectively.

Realistically, in order to determine the exact amount of delay
time, we have to measure the time profiles at the detector of both
the KeV/MeV and of the TeV bursts, respectively. If the time delay
between the KeV/MeV bursts and TeV bursts is longer than several
seconds, the ground based TeV telescopes are capable to catch the
burst. A good coordinate effort between the SWIFT satellite and
the ground based TeV telescopes can easily make this measurement
possible \citep{We03}.

Observations of the time delay in gamma ray bursts have been
proposed up to now mainly as tests of quantum gravity effects. In
a pioneering work \citet{Am98} suggested that
quantum-gravitational effects could induce a deformed dispersion
relation for photons of the form $c^{2}$ $\vec{p}^2=E^{2}\left[
1+f\left( E/E_{QG}\right) \right] $. By representing $\ f$ in a
form of a power series, the energy dependent velocity of the speed
of light can be represented as $v\approx c\left( 1\pm
E/E_{QG}\right) $. This type of velocity dispersion results from a
picture of the vacuum as a quantum gravitational 'medium', which
responds differently to the propagation of particles of different
energies. This is analogous to propagation through a conventional
medium, such as an electromagnetic plasma. The gravitational
'medium' is generally believed to contain microscopic quantum
fluctuations, which may occur on scale sizes of order of the
Planck length. In this approach the vacuum is viewed as a
non-trivial medium containing 'foamy' quantum-gravity
fluctuations. The nature of this foamy vacuum may be visualized,
for example, by imagining processes that include the pair creation
of virtual black holes. The light propagation in the semiclassical
space-time that emerges in canonical quantum gravity in the loop
representation was considered in \cite{Ga99}. In such a picture
space-time exhibits a polymer-like structure at micro-scales, and
departures from the perfect non-dispersiveness of an ordinary
vacuum naturally occur. Maxwell equations are modified due to the
quantum gravity, and non-vanishing corrections to the
electromagnetic field equations appear that depend on the helicity
of the propagating waves. These effects could lead to constraints
on the discrete nature of the quantum space-time from the study of
gamma-ray bursts at different energies.

In the present paper we have considered a different class of
effects, which could generate a time delay $\Delta t$ of the high
energy photons, emitted during the gamma ray bursts, namely the
possibility that the extra-dimensions of the space-time may modify
the speed of light in vacuum. This is mainly due to the dependence
of the fine structure constant $\alpha $ on the extra-dimensions.
The effects related to extra-dimensions in the variation of
$\alpha $ are much stronger than the quantum gravity effects. In
models with extra-dimensions, the speed of light-photon energy
dependence can be obtained exactly, by considering the background
gravitational field and the scalar and Yang-Mills type gauge
fields in extra-dimensions. In quantum-gravity models this
dependence is modelled more or less phenomenologically. On the
other hand, the observation of the time delay of the photons in
GRBs could provide some astrophysical tests for the confirmation
of the existence of the extra-dimensions of the space-time.

We have analyzed in detail the time delay for models with both
compact and non-compact extra-dimensions, deriving some explicit
expressions for $\Delta t$. In the case of Kaluza-Klein type
theories the compactification of extra dimensions provides a
natural framework for the variation of the fine structure constant
and for the speed of light. The basic energy scale for this model
is of the order of $10^{19}$ GeV, which for Einstein-Yang-Mills
type models gives a delay that can be as high as $\Delta t=10^2$ s
for $1$ TeV photons emitted at a redshift of $z=5$. In these
models the variation in the fine structure constant is dominated
by the effects of the extra-dimensions, and the quantum effects
can be neglected.

Since in models with large extra dimensions the energy scale can
be reduced significantly, as a function of the size of $v$, much
higher time delays are expected, which could be of the order of
$10^5$ s for $1$ TeV photons. There is a strong model-dependence
of the Kaluza-Klein type time delay expressions, the results
depending on the initial (multi-dimensional) field structure. For
a pure Einstein gravity in higher dimensions the photon time delay
has extremely small values, which make it extremely difficult to
detect.

For the detection of quantum gravity or extra-dimensional effects
gamma ray bursts offer the most reliable high energy photon
sources, located at cosmological distances. The BATSE data have
already been used to extract valuable information on the quantum
gravity energy scale \citep{El03}. However, these data are
restricted to a low energy range, of the order of $30-300$ keV. In
this range other concurrent physical processes, like Compton
scattering, could reduce the effect. In order to obtain more
reliable data a significant increase in the detected photon energy
is necessary. The detection of quantum gravitational and
extra-dimensional features would also require the correlation of
GRB redshifts with the temporal and energetic signatures.

The discovery of the linear polarization of the $\gamma $-rays
from the GRBs, with the estimated degree of polarization of $80\pm
20\%$, very close to the absolute maximum of $100\%$, provides an
other test of quantum gravity effects \citep{Mi03}. If the effects
of quantum gravity are linearly proportional to the ratio
$E/E_{QG}$, then the polarization of photons with energies of
about $0.1$ MeV should be completely random, contrary to the
observations. Consequently, quantum gravity effects act with a
power greater than one. The linear polarization of $\gamma $ rays
also allow to test the birefringence property of the quantum
vacuum, as suggested by the quantum gravity in loop representation
\citep{Ga99}. Due to this property, two photons with opposite
states of helicity have different group velocities
\citep{GlKo01,Mi03}. A significant rotation of the plane of
polarization of a linearly polarized photon must occur long before
any difference in time of arrival is even measurable. If a
rotation, if present, is below a certain bound, one can obtain a
general bound on the model parameters characterizing the effect
\citep{GlKo01}. By analyzing the presence of linear polarization
in the optical and ultraviolet spectrum of some distant sources,
the limit $\chi <5\times 10^{-5}$ has been found for the
dimensionless parameter $\chi $ that characterizes both parity
non-conservation and violation of Lorentz invariance
\citep{GlKo01}. This upper bound on $\chi $ induces a time delay
of the order of $10^{-9}$ s, which is beyond the possibility of
observation. However, for cosmological GRBs, located at a distance
of around $10^{10}$ light years, the quadratic birefringence of
quantum space-time could be tested by polarization measurements of
photons with energies greater than $100$ MeV \citep{Mi03}. On the
other hand, in models with extra-dimensions, there is a third
order dependence on energy of the speed of light, which could also
be tested by using polarimetry of $\gamma $-rays from cosmological
sources. The combinations of both approaches based on time-delay
measurements and polarimetry could provide significant constraints
on quantum gravity and multi-dimensional models.

Ground-based atmospheric Cherenkov telescopes offer a unique
opportunity for the observation of the delayed TeV components of
gamma-ray bursts. In the past few years such telescopes using the
imaging technique have proved to be remarkably sensitive for the
detection of sources with hard gamma ray spectra \citep{We03}.
Because of the very large collection areas associated with these
telescopes ($>50,000$ m$^2$), they are particularly sensitive to
the detection of transients, e. g. from AGN such as Markarian 421
\citep{Ga96}. These telescopes are more sensitive than the all-sky
viewing ground-level particle detectors such as Milagro, the Tibet
array or Argo, but have limited fields of view. However, they have
been used in attempts to detect TeV emission from classical gamma
ray bursts \citep{Co97} and from primordial black holes
\citep{Co98}. Although the detectable fluence can be as little as
$10^{-8}$ ergs/sec, no detections have been reported yet. However,
in no instance has a gamma-ray burst been reported within the
field of view of an operating imaging atmospheric Cherenkov
telescope. Reported observations have been limited by the slew
time of the telescope, the uncertainty in the initial source
position and the limited time of operation.

Therefore, the detection of the time delay between Tev and GeV/MeV
photons from GRBs could represent a new possibility for the study
and understanding of some fundamental aspects of the physical laws
governing our universe.

\section*{Acknowledgments}

This work is supported by a RGC grant of the Hong Kong Government.
The authors would like to thank Prof. T. C. Weekes, Prof. Z. G.
Dai, Dr. Y. F. Huang, Dr. X. Y. Wang  and to the anonymous referee
for suggestions, which significantly improved the manuscript.

\begin{figure}
\plotone{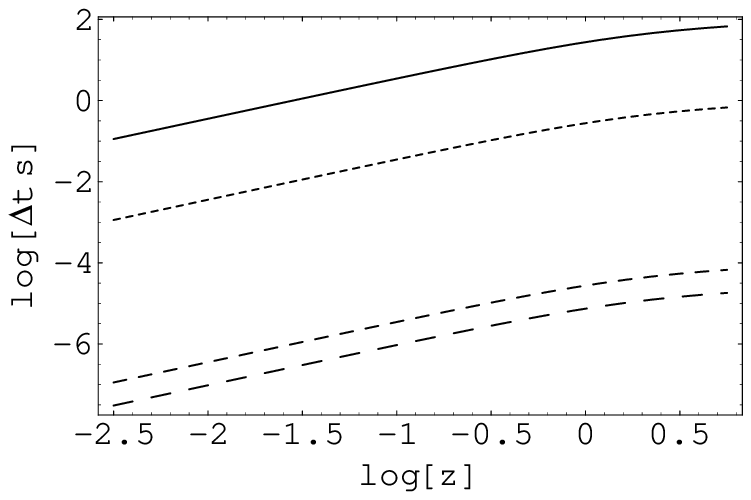}
\caption{ Variation of the photon time delay as a function of the
redshift $z$ (in a logarithmic scale) in the compactified
Kaluza-Klein model, with Einstein-Yang-Mills type action ($\beta
=1$), for $\gamma =1$ (corresponding to a fundamental energy scale
$E_K=1.2\times 10^{19}$ GeV), and for different photon energy
values: $E_1=1$ TeV, $E_2=1$ eV (solid curve),
 $E_1=10$ GeV, $E_2=1$ MeV (dotted curve), $E_1=1$ MeV, $E_2=1$ eV (short dashed curve)
 and $E_1=300$ keV, $E_2=30$ keV (long dashed curve). For the mass and dark energy parameters
 we have used the values $\Omega _{M}= 0.3$ and $\Omega _{\Lambda }= 0.7$, respectively.
 }
\label{FIG1}
\end{figure}

\begin{figure}
\plotone{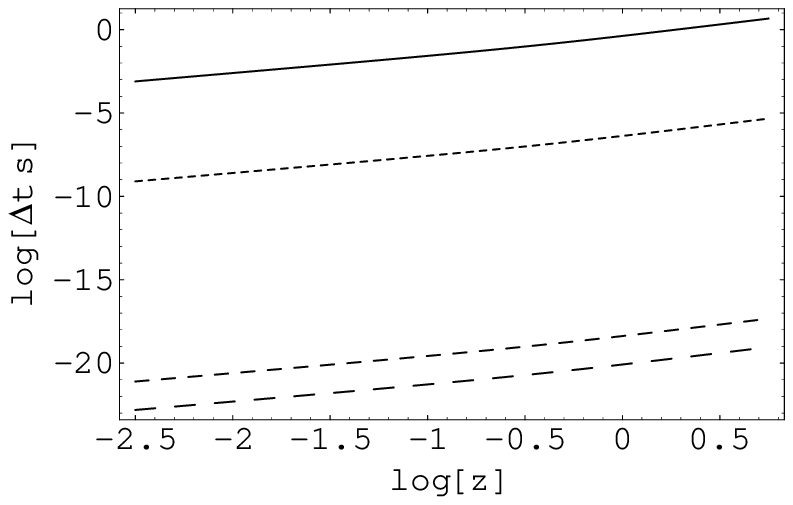}
\caption{ Variation of the photon time delay as a function of the
redshift $z$ (in a logarithmic scale) in the compactified
Kaluza-Klein model with pure Einstein type action ($\beta =3$),
for $\gamma =10^{10}$ (corresponding to a fundamental energy scale
 $E_K=1.2\times 10^9$ GeV), and for different photon energy values: $E_1=1$ TeV, $E_2=1$ eV
(solid curve), $E_1=10$ GeV, $E_2=1$ MeV (dotted curve), $E_1=1$
MeV, $E_2=1$ eV (short dashed curve)
 and $E_1=300$ keV, $E_2=30$ keV (long dashed curve). For the mass and dark energy parameters
 we have used the values $\Omega _{M}= 0.3$ and $\Omega _{\Lambda }= 0.7$, respectively.
 }
\label{FIG2}
\end{figure}

\begin{figure}
\plotone{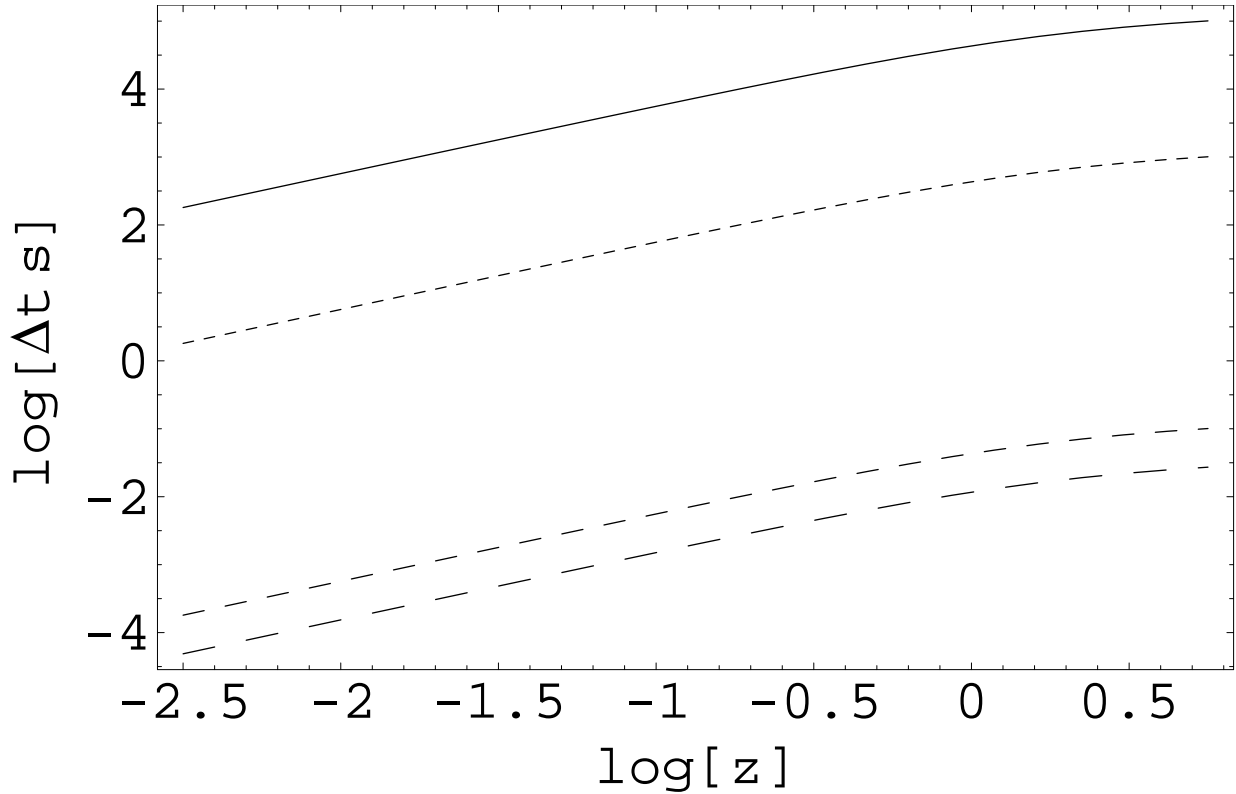}
\caption{ Variation of the photon time delay as a function of the
redshift $z$ (in a logarithmic scale) in cosmological models with
large extra-dimensions for a fundamental energy scale $E_F=7\times
10^{15}$ GeV and for different photon energy values: $E_1=1$ TeV,
$E_2=1$ eV (solid curve), $E_1=10$ GeV, $E_2=1$ MeV (dotted
curve), $E_1=1$ MeV, $E_2=1$ eV (short dashed curve)
 and $E_1=300$ keV, $E_2=30$ keV (long dashed curve). For the mass, dark energy and dark radiation parameters
 we have used the values $\Omega _{M}=0.3$, $\Omega _{\Lambda }=0.68$ and $\Omega _U=0.02$, respectively.
 }
\label{FIG3}
\end{figure}

\begin{figure}
\plotone{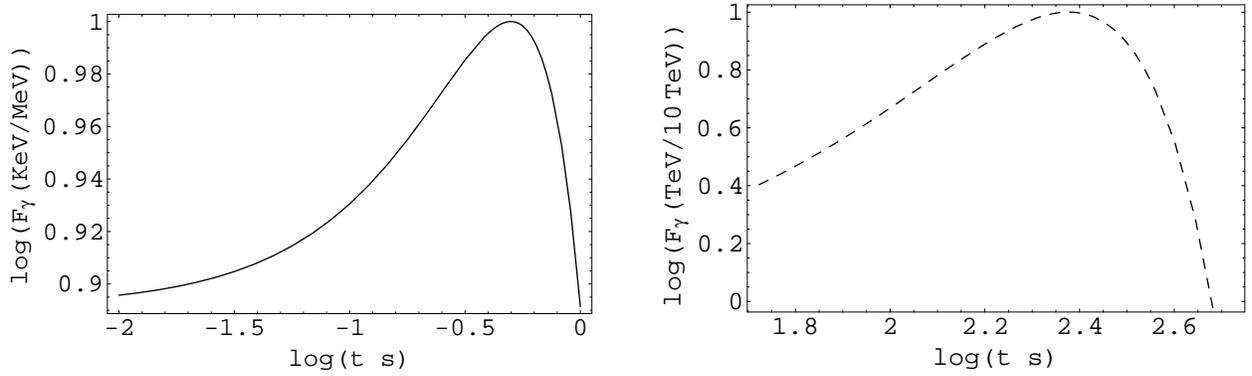}
\caption{Comparison, in arbitrary units, of the initial KeV/MeV
time profile of the GRB emission occurring at a redshift $z=3$
(assumed to have a Gaussian form), with a duration of $\tau =1$ s
(solid curve), and the TeV time profile at the detector, modified
due to the presence of multi-dimensional and/or quantum gravity
effects (dashed curve). Both time profiles have been normalized to
$1$. For the photon time delay we have adopted the linear model,
with the fundamental energy scale $E_K=1.2\times 10^{19}$ GeV. The
photon energies are in the range E$\in \left(1,10\right) $ TeV.
For the mass and dark energy parameters we have used the values
$\Omega _{M}=0.3$ and $\Omega _{\Lambda }=0.7$, respectively.}
\label{FIG4}
\end{figure}


\begin{thebibliography}{99}

\bibitem[Aharonanian et al.(1997)]{Ah97} Aharonian, F. A.  et al., 1997, Astron. Astrophys., 327, L5

\bibitem[Amelino-Camelia et al.(1998)]{Am98} Amelino-Camelia, G., Ellis, J., Mavromatos,  N. E.,
Nanopoulos, D. V. \& Sarkar, S., 1998, Nature, 393, 763

\bibitem[Atkins et al.(2000)]{At00} Atkins, R. et al., 2000, Astrophys. J., 533, L119

\bibitem[Atoyan and Aharonian(1996)]{At96} Atoyan,  A. M. \& Aharonian, F. A., 1996,  Mon. Not. R. Astron.
Soc., 278, 525

\bibitem[Biller et al.(1999)]{Bi} Biller,  S. D. et al., 1999, Phys. Rev. Lett., 83, 2108

\bibitem[Boos et al.(2002)]{Bo02} Boos, E. E., Kubyshin, Y. A., Smolyakov, M. N. \&
Volobuev, I. P., 2002, Class. Quant. Grav., 19, 4591

\bibitem[Chen et al.(2001a)]{Chph} Chen, C. M., Harko, T. \& Mak, M. K., 2001a, Phys. Rev. D, 64, 044013

\bibitem[Chen et al.(2001b)]{Ch2} Chen, C. M., Harko, T. \& Mak, M. K., 2001b, Phys. Rev. D, 64, 124017

\bibitem[Chen et al.(2002)]{Chnu}  Chen, C. M.,  Harko, T., Kao, W. F. \& Mak, M. K., 2002, Nuclear Physics B, 636, 159

\bibitem[Chen et al.(2003)]{Ch03} Chen, C. M., Harko, T., Kao, W. F.  \& Mak, M. K., 2003, JCAP 0311, 005.

\bibitem[Cheng and Cheng(1996)]{Ch96} Cheng, L. X. \&  Cheng, K. S., 1996,  Astrophys. J., 459, L79


\bibitem[Cheng and Lu(2001)]{ChLu01}Cheng, K. S. \&  Lu, T., 2001,
Chin. J. Astron. Astrophys., 1, 1
\bibitem[Connaughton et al.(1997)]{Co97} Connaughton, V. et al., 1997, Astrophys. J., 479, 859

\bibitem[Connaughton et al.(1998)]{Co98} Connaughton, V. et al., 1998, Astropart. Phys., 8, 179

\bibitem[Dabrowski et al.(2002)]{Da02a} Dabrowski, M. P., Godlowski, W. \& Szydlowski, M., 2002, astro-ph/0212100

\bibitem[Dai and Lu(2002)]{Da02} Dai, Z. G. \& Lu, T., 2002, Astrophys. J., 580, 1013

\bibitem[Dirac(1937)]{Di38} Dirac, P. A. M., 1937, Nature, 139, 323

\bibitem[Dirac(1938)]{Di39} Dirac, P. A. M., 1938, Proc. R. Soc. London A, 165, 198

\bibitem[Dirac(1979)]{Di79} Dirac, P. A. M., 1979, Proc. R. Soc. London A, 365, 19

\bibitem[Ellis et al.(2000a)]{El00c} Ellis, J., Mavromatos, N. E. \& Nanopoulos, D. V., 2000a,  Phys.
Rev. D, 61, 027503

\bibitem[Ellis et al.(2000b)]{El00} Ellis, J., Mavromatos, N. E. \& Nanopoulos, D. V., 2000b, Gen. Rel.
Grav., 32, 127

\bibitem[Ellis et al.(2000c)]{E101} Ellis, J., Mavromatos, N. E. \& Nanopoulos, D. V., 2000c, Phys. Rev.
D, 62, 084019

\bibitem[Ellis et al.(2000d)]{Ma00} Ellis, J., Farakos, K., Mavromatos, N. E., Mitsou,  V. \&
Nanopoulos, D. V., 2000d, Astrophys. J., 535, 139

\bibitem[Ellis et al.(2003a)]{ElM03} Ellis,  J., Mavromatos,  N. E. \& Sakharov, A. S.,
2003a, astro-ph/0308403

\bibitem[Ellis et al.(2003b)]{El03} Ellis, J., Mavromatos, N. E.,  Nanopoulos, D. V. \&
Sakharov, A. S., 2003b, Astron. Astrophys., 402, 409

\bibitem[Freedman et al.(2001)]{Fr01}Freedman, W. L. et al., 2001, Astrophys. J., 553, 47

\bibitem[Gaidos et al.(1996)]{Ga96} Gaidos,  J. A. et al., 1996, Nature, 383, 319

\bibitem[Gambini and Pullin(1999)]{Ga99} Gambini R. \& Pullin, J., 1999, Phys. Rev. D, 59, 124021

\bibitem[Gleiser and Kozameh(2001)]{GlKo01} Gleiser R. J. \&
Kozameh, C. N., 2001, Phys. Rev. D, 64, 083007



\bibitem[Harko and Mak(2003)]{Ha03} Harko, T. \&  Mak, M. K., 2003, Class. Quantum Grav.,  20, 407

\bibitem[Hillas et al.(1998)]{Hi98} Hillas  A. H. et al, 1998, Astrophys. J., 503, 744

\bibitem[Horava and Witten(1996)]{HoWi96} Horava, P. \& Witten, E., 1996, Nucl. Phys. B, 475, 94

\bibitem[Hurley(1994)]{Hu94} Hurley, K., 1994, Nature, 371, 652

\bibitem[Jacobson et al.(2003)]{Ja03} Jacobson,  T.,  Liberati, S. \&  Mattingly, D.,
2003a, Nature, 424, 1019

\bibitem[Jacobson et al.(2003)]{Jal03} Jacobson, T., Liberati, S., Mattingly,  D. \&
Stecker, F. W., 2003, astro-ph/0309681

\bibitem[Kaaret(1999)]{Ka99} Kaaret,  P., 1999, Astron. Astrophys., 345, L32

\bibitem[Kaluza(1921)]{Kl19}  Kaluza, T., 1921, Sitzungsberichte Preussische Akademie der
Wissenschaften, 96, 69

\bibitem[Katz(1994)]{Ka94} Katz, J. I., 1994, Astrophys. J., 432, L27

\bibitem[Klein(1926)]{Kl26} Klein, O., 1926, Zeitschrift fur Physik, 37, 895


\bibitem[Krennrich et al.(2001)]{Kr01} Krennrich, F. et al., 2001, Astrophys. J., 560, L45


\bibitem[Langacker et al.(2002)]{La02} Langacker,  P. , Segre, G. \& Strassler, M. J., 2002, Phys. Lett.
B, 528, 121

\bibitem[Loren-Aguilar et al.(2003)]{Lo03} Loren-Aguilar,  P., Garcia-Berro, E., Isern,  J. \&
Kubyshin, Yu. A., 2003, Class. Quant. Grav., 20, 3885

\bibitem[Ma(1990a)]{Ma90} Ma, G. W., 1990a, Phys. Lett. A, 146, 375

\bibitem[Ma(1990b)]{Ma90a} Ma, G. W., 1990b, Phys. Lett. A, 143, 183

\bibitem[Madau and Phinney(1996)]{Ma96} Madau, P. \& Phinney, E. S., 1996, Astrophys. J.,
456, 124

\bibitem[Maeda and Wands(2000)]{MW00} Maeda, K. \& Wands, D., 2000,  Phys. Rev. D, 62, 124009

\bibitem[Magueijo(2003)]{Ma03} Magueijo,  J. , 2003, Reports on Progress in Physics, 66, 2025

\bibitem[Mak and Harko(1999)]{MaHa99} Mak, M. K. \& Harko, T., 1999, Class. Quantum Grav., 16, 4085

\bibitem[Meszaros and Rees(1994)]{MeRe94} Meszaros, P. \& Rees,
M. J., 1994, Mon. Not. Roy. Astron. Soc., 269, L41


\bibitem[Mitrofanov(2003)]{Mi03} Mitrofanov, I. G., 2003, Nature,
426, 139

\bibitem[Murphy et al.(2001)]{We2} Murphy, M. T. et al., 2001, Mon. Not. Roy. Astron. Soc., 327, 1208

\bibitem[Murphy et al.(2003)]{We3} Murphy,  M. T., Webb, J. K. \& Flambaum, V. V., 2003, Mon. Not. Roy.
Astron. Soc., 345, 609


\bibitem[Norris et al.(1999)] {No} Norris, J. P., Bonnell, J. T., Marani, G. F. \&
Scargle, J. D., 1999, astro-ph/9912136

\bibitem[Overduin and Wesson(1997)]{Ov97} Overduin, J. M. \& Wesson, P. S., 1997, Physics Reports, 283, 303

\bibitem[Peebles and Ratra(2003)]{Pe03} Peebles, P. J. E. \& Ratra, Bharat, 2003, Rev. Mod. Phys., 75, 559


\bibitem[Plaga(1995)]{Pl95} Plaga, R., 1995, Nature, 374, 430

\bibitem[Poirier et al.(2003)]{Po03} Poirier, J., D'Andrea,  C. , Fragile, P. C., Gress,  J., Mathews, G. J. \&
Race, D., 2003, Phys. Rev. D, 67, 042001

\bibitem[Randall and Sundrum(1999a)]{Ra98} Randall, L. \& Sundrum, R., 1999a, Phys. Rev. Lett., 83, 3370

\bibitem[Randall and Sundrum(1999b)]{Ra98a} Randall, L. \& Sundrum, R., 1999b, Phys. Rev. Lett., 83, 4690



\bibitem[Sasaki et al.(2000)]{SaShMa00} Sasaki, M., Shiromizu, T. \& Maeda, K., 2000, Phys. Rev. D, 62,
024008

\bibitem[Shiromizu et al.(2000)]{SMS00}Shiromizu, T., Maeda, K. \& Sasaki, M., 2002, Phys. Rev. D 62
(2000), 024012

\bibitem[Sommer et al.(1994)]{So94} Sommer, M. et al., 1994, Astrophys. J., 422, L63



\bibitem[Stecker(2003)]{St03} Stecker, F. W., 2003, Astropart. Phys., 20, 85


\bibitem[Stecker et al.(1996)]{St96}  Stecker, F. W., de Jager, O. C. \& Salamon, M. H., 1996, Astrophys. J.
473, L75




\bibitem[Uzan(2003)]{Uz03} Uzan  J. P. , 2003, Rev . Mod. Phys., 75, 403


\bibitem[Wang et al.(2004)]{WaCh03} Wang, X. Y., Cheng, K. S., Dai, Z. G. \& Lu, T.,
2004, Astrophys. J., in press; astro-ph/0311601

\bibitem[Webb et al.(1999)]{We1}  Webb, J. K. et al., 1999, Phys. Rev. Lett., 82, 884

\bibitem[Weekes et al.(2002)]{We02} Weekes, T. C. et al., 2002, Astropart. Phys., 17, 221

\bibitem[Weekes(2003)]{We03} Weekes, T. C., 2003, Plenary talk at the 28th I. C.
R. C. (Tsukuba, Japan), in press



\bibitem[Witten(1996)]{Wi96} Witten E., 1996, Nucl. Phys. B, 460, 335

\bibitem[Zhang and Meszaros(2003)]{ZhMe03} Zhang, B. \& Meszaros, P., 2003,
Int. J. Mod. Phys. A, in press; astro-ph/0311321

\end{thebibliography}
\end{document}